\documentstyle[jkas,times]{article}

\month{0} \year{2011} \volume{0} \issueno{0}

\runningauthor{M. CHOI AND J.-E. LEE} 
\runningtitle{NGC 1333 IRAS 4B RADIO IMAGING} 

\begin{document}

\title{RADIO IMAGING OF THE NGC 1333 IRAS 4B REGION} 
\author{Minho Choi$^{1,2}$ and Jeong-Eun Lee$^3$} 
\address{$^1$ Korea Astronomy and Space Science Institute,
              776 Daedeokdaero, Yuseong, Daejeon 305-348 \\
              {\it E-mail : minho@kasi.re.kr}}
\address{$^2$ University of Science and Technology,
              217 Gajeongro, Yuseong, Daejeon 305-350}
\address{$^3$ Department of Astronomy and Space Science,
              Kyung Hee University, Yongin, Gyeonggi 446-701}
\offprints{M. Choi}

\abstract{
The NGC 1333 IRAS 4B region was observed in the 6.9 mm and 1.3 cm continuum
with an angular resolution of about 0.4 arcseconds.
IRAS 4BI was detected in both bands,
and BII was detected in the 6.9 mm continuum only.
The 1.3 cm source of BI seems to be a disk-like flattened structure
with a size of about 50 AU.
IRAS 4BI does not show any sign of multiplicity.
Examinations of archival infrared images show
that the dominating emission feature in this region
is a bright peak in the southern outflow driven by BI,
corresponding to the molecular hydrogen emission source HL 9a.
Both BI and BII are undetectable in the mid-IR bands.
The upper limit on the far-IR flux of IRAS 4BII suggests
that it may be a very low luminosity young stellar object.}

          
\maketitle

\section{INTRODUCTION}

NGC 1333 IRAS 4B is a Class 0 young stellar object
in the Perseus molecular cloud at a distance of 235 pc from the Sun
(Sandell et al. 1991; Hirota et al. 2008; Enoch et al. 2009).
Radio and infrared observations revealed
various star forming activities in this region,
such as molecular outflows and H$_2$O masers
(Blake et al. 1995; Hodapp \& Ladd 1995; Choi et al. 1999, 2004, 2006;
Rodr{\'\i}guez et al. 1999, 2002; Choi 2001; Reipurth et al. 2002;
Park \& Choi 2007).
Interferometric observations revealed
that there are two compact continuum sources, BI and BII,
with an angular separation of 11$''$ (Looney et al. 2000; Choi 2001).
BI is a protostellar object,
and most of the star forming activities in this region
can be attributed to BI.
By contrast, BII has been detected
in the millimeter and submillimeter wavelength bands only,
and its nature is less clear.

Lay et al. (1995) performed a single-baseline submillimeter interferometry
with a resolution of $\sim$0\farcs5
and suggested that IRAS 4BI is a multiple system.
However, later studies could not confirm the multiplicity of BI
(Looney et al. 2000; Reipurth et al. 2002).
To settle this issue, it is necessary to make sensitive observations
with a resolution higher than 0\farcs5.

While IRAS 4BII has received relatively less attention,
it is probably an interesting object
in understanding low-mass star formation processes,
because its nondetection in infrared
suggests that it is an unusually cold object.
The lack of detectable star forming activity is also intriguing.

In this paper, we present the results of
our observations of the NGC 1333 IRAS 4B region
in the 6.9 mm and 1.3 cm continuum with the Very Large Array (VLA)
of the National Radio Astronomy Observatory.
We describe our radio continuum observations in Section 2.
In Section 3, we report the results of the continuum imaging.
In Section 4, we discuss the star-forming activities in the IRAS 4B region.

\section{OBSERVATIONS AND DATA}

\subsection{Q-band Observations}

The NGC 1333 IRAS 4B region was observed using VLA
in the standard Q-band continuum mode (43.3 GHz or $\lambda$ = 6.9 mm).
Twenty-five antennas were used
in the C-array configuration on 2004 March 2.
The phase-tracking center was
$\alpha_{2000}$ = 03$^{\rm h}$29$^{\rm m}$12\fs988,
$\delta_{2000}$ = 31\arcdeg13$'$08\farcs10,
which is $\sim$2$''$ east and $\sim$1$''$ north of IRAS 4BII.

The phase was determined
by observing the nearby quasar 0336+323 (PKS 0333+321).
The flux was calibrated
by setting the flux density of the quasar 0713+438 (QSO B0710+439) to 0.20 Jy,
which is the flux density measured within a day of our observations
(VLA Calibrator Flux Density Database%
\footnote{See http://aips2.nrao.edu/vla/calflux.html.}).
Comparison of the amplitude gave a flux density of 1.36 Jy for 0336+323.
To avoid the degradation of sensitivity owing to pointing errors,
pointing was referenced
by observing the calibrators at the X band ($\lambda$ = 3.6 cm).
This referenced pointing was performed
about once an hour and just before observing the flux calibrator.

Maps were made using a CLEAN algorithm.
With a natural weighting,
the 6.9 mm continuum data produced a synthesized beam
of $\sim$0\farcs5 in full-width at half-maximum (FWHM).

\begin{figure*}[!t]
\centering
\epsfxsize=17cm 
\epsfbox{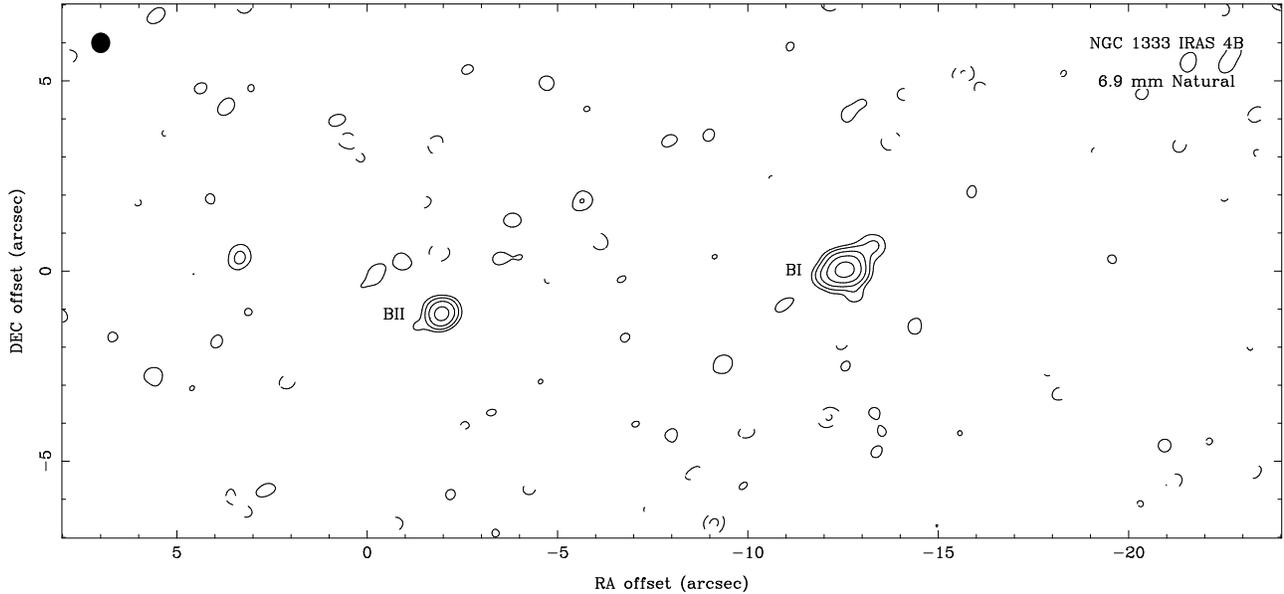} 
\caption{
Map of the $\lambda$ = 6.9 mm continuum emission
toward the NGC 1333 IRAS 4B region,
made with a natural weighting.
The contour levels are 1, 2, 4, 8, and 16 $\times$ 0.07 mJy beam$^{-1}$.
Dashed contours are for negative levels.
The rms noise is 0.024 mJy beam$^{-1}$.
Shown in the top left-hand corner is the synthesized beam:
FWHM = 0\farcs53 $\times$ 0\farcs51 with PA = --1$^\circ$.
The coordinates are position offsets
relative to the phase tracking center of the Q-band observations.}
\end{figure*}

\begin{table*}[!t]
\caption{NGC 1333 IRAS 4B Continuum Source Parameters}
\begin{center}
\begin{tabular}{lcccccccc}
\hline \hline
& \multicolumn{2}{c}{Peak Position$^a$}
&& \multicolumn{2}{c}{6.9 mm Flux Density$^b$}
&& \multicolumn{2}{c}{1.3 cm Flux Density$^b$} \\
\cline{2-3} \cline{5-6} \cline{8-9}
Source & $\alpha_{\rm J2000.0}$ & $\delta_{\rm J2000.0}$
&& Peak & Total && Peak & Total \\
\hline
BI  & 03 29 12.009 & 31 13 08.14 && 1.77 $\pm$ 0.03 & 3.08 $\pm$ 0.10
    && 0.362 $\pm$ 0.019 & 0.47 $\pm$ 0.05 \\
BII & 03 29 12.835 & 31 13 06.98 && 0.77 $\pm$ 0.02 & 0.89 $\pm$ 0.05
    && \ldots & \ldots \\
\hline
\end{tabular}
\parbox{149mm}{\medskip
$^a$ Source positions from the robust-weight 6.9 mm map (Figure 2).
     The 1.3 cm position of BI agrees with the 6.9 mm position
     within 0\farcs09.
     Units of right ascension are hours, minutes, and seconds,
     and units of declination are degrees, arcminutes, and arcseconds. \\
$^b$ Flux densities are in mJy beam$^{-1}$ or mJy,
     corrected for the primary beam response.}
\end{center}
\end{table*}

\subsection{K-band Data}

The NGC 1333 IRAS 4 region was observed using VLA
in the $\lambda$ = 1.3 cm continuum.
Details of the observations and the results for the IRAS 4A region
were presented in Choi et al. (2010, 2011).
The observations were made
in the standard K-band continuum mode (22.5 GHz)
in the B-array configuration.
With a natural weighting,
the 1.3 cm continuum data produced a synthesized beam
of FWHM $\approx$ 0\farcs4.

\section{RESULTS}

Figure 1 shows the 6.9 mm continuum map
toward the NGC 1333 IRAS 4B region.
IRAS 4BI and BII were clearly detected.
Table 1 lists the continuum source parameters.
Figure 2 shows a map of the BI region with a smaller beam.
Figure 3 shows the same region in the 1.3 cm continuum.
IRAS 4BII was not detected in the 1.3 cm continuum.

\begin{figure}[!t]
\centering
\epsfxsize=8cm 
\epsfbox{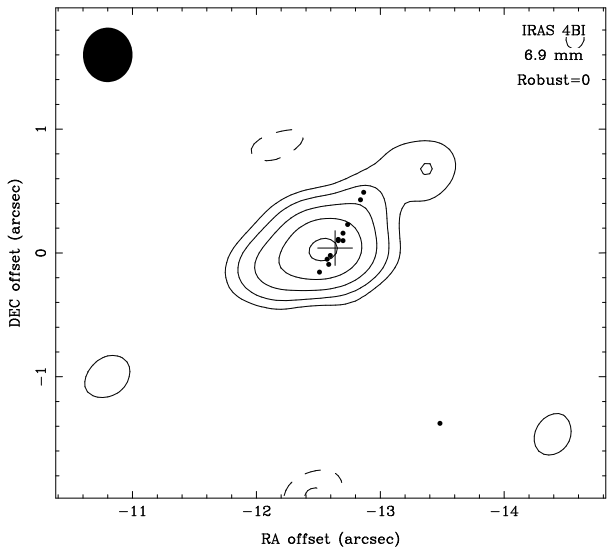} 
\caption{
Map of the $\lambda$ = 6.9 mm continuum emission
toward the IRAS 4BI region,
made with a robust weighting.
The contour levels are 1, 2, 4, 8, and 16 $\times$ 0.07 mJy beam$^{-1}$.
The rms noise is 0.024 mJy beam$^{-1}$.
Shown in the top left-hand corner is the synthesized beam:
FWHM = 0\farcs45 $\times$ 0\farcs40 with PA = --1$^\circ$.
Solid dots:
H$_2$O maser sources (Rodr{\'\i}guez et al. 2002; Park \& Choi 2007).
Plus sign:
the 3.6 cm continuum source (Reipurth et al. 2002).}
\end{figure}

\begin{figure}[!t]
\centering
\epsfxsize=8cm 
\epsfbox{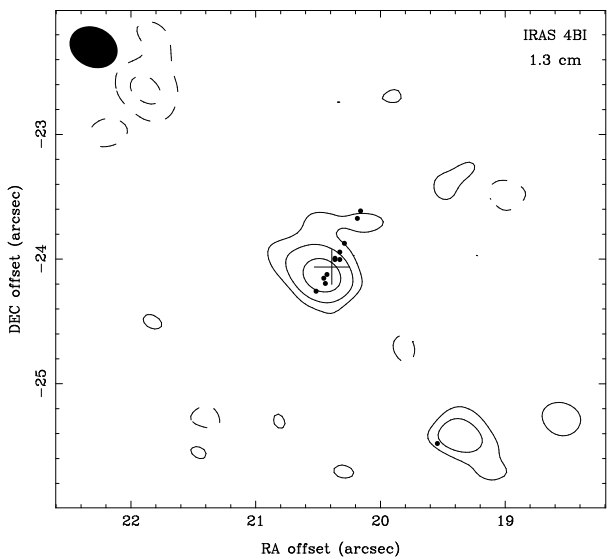} 
\caption{
Map of the $\lambda$ = 1.3 cm continuum toward the IRAS 4BI region.
The contour levels are 1, 2, and 4 $\times$ 0.05 mJy beam$^{-1}$.
The rms noise is 0.016 mJy beam$^{-1}$.
Shown in the top left-hand corner is the synthesized beam:
FWHM = 0\farcs40 $\times$ 0\farcs33 with PA = 64$^\circ$.
The coordinates are position offsets
relative to the phase tracking center of the K-band observations.}
\end{figure}

In the 6.9 mm map (Figure 2),
IRAS 4BI seems to be elongated in the northwest-southeast direction.
An elliptical Gaussian fit gives
a deconvolved source size of FWHM = 0\farcs56 $\times$ 0\farcs15
with a position angle (PA) of --73$^\circ$.
Comparison with the distribution of the H$_2$O maser sources
(Rodr{\'\i}guez et al. 2002; Park \& Choi 2007) suggests
that the 6.9 mm continuum source structure is affected by the outflow,
at least in the northwestern part of the source.
Possible explanations include
shock-heated dusts around the outflow
and free-free emission from ionized outflowing gas.

In the 1.3 cm map (Figure 3),
IRAS 4BI seems to be elongated in the northeast-southwest direction.
An elliptical Gaussian fit gives
FWHM = 0\farcs20 $\times$ 0\farcs10 with PA = 50$^\circ$.
The size corresponds to 47 $\times$ 24 AU$^2$ at a distance of 235 pc.
Since the elongation is nearly perpendicular
to the direction of the H$_2$O outflow (PA $\approx$ 151$^\circ$),
the 1.3 cm continuum may be tracing
either a circumstellar disk or a flattened protostellar envelope (pseudodisk),
or probably a combination of them.

The 6.9 mm continuum source of IRAS 4BII (Figure 1)
is essentially unresolved.
BII is very compact and has never been clearly resolved
in interferometric images (e.g., Looney et al. 2000;
note that IRAS 4BII was referred to as IRAS 4C by Looney et al.).

The total flux densities from our data (Table 1)
are smaller than those given by Di Francesco et al. (2001)
by a factor of $\sim$2.5.
One possible reason for the difference is
that our observations were made in more extended array configurations.
Another reason could be that the flux calibrator in our observations is different from that used by Di Francesco et al. (2001).

\section{DISCUSSION}

\subsection{IRAS 4BI}

\subsubsection{Source Structure}

The single-baseline submillimeter interferometric observations
of Lay et al. (1995) suggested
that the source structure of IRAS 4BI cannot be explained as a single source.
Even a binary model could not fit the data well,
and they suggested that BI is a triple or higher-order multiple system.
Considering their resolution and the clear dips in the visibility curve,
BI was expected to consist of multiple components
of comparable brightness with a separation larger than 0\farcs5.
In later studies with resolutions higher than 1$''$, however,
the image of IRAS 4BI was always found to be dominated by a single peak
(Looney et al. 2000; Reipurth et al. 2002; Choi et al. 2007;
J{\o}rgensen et al. 2007; J{\o}rgensen \& van Dishoeck 2010).

Our 6.9 mm continuum map (Figure 2) shows a weak peak
located about 1$''$ northwest of IRS 4BI.
This peak is probably not a real source,
considering that the map shows some negative peaks
at comparable (absolute) intensities
and that the 2.7 mm map of Looney et al. (2000)
does not show a corresponding peak.
Even if it is real,
it could not have caused the interference pattern seen by Lay et al. (1995)
because it is weaker than the BI main peak by a factor of $\sim$9.
The elongated shape of the source (Section 3)
could have caused some interference pattern.
Another possible source of confusion is BII,
because it is located in the primary beam of Lay et al.
and its existence was not known at the time of their analysis.

\subsubsection{Continuum Spectrum and Mass}

Figure 4 shows the spectral energy distribution (SED) of IRAS 4BI.
(For the SED and mass estimates,
we use the flux densities given by Di Francesco et al. (2001),
7.8 mJy at 6.9 mm and 1.2 mJy at 1.3 cm,
because the values in Table 1 might suffer from missing flux.)
The steep slope in the (1.3 cm, 2.7 mm) section suggests
that the millimeter flux comes from thermal emission of dust.
The SED in this section can be described using a power-law form,
$F \propto \nu^\alpha$, where $F$ is the flux density,
$\nu$ is the frequency, and $\alpha$ is the spectral index.
The best-fit power-law spectrum gives $\alpha$ = 3.3 $\pm$ 0.2.
The 3.5--6.2 cm flux may be mostly
from free--free emission of a thermal radio jet.
The free-free emission seems to be optically thick at 6.2 cm.

\begin{figure}[!t]
\centering \epsfxsize=8cm 
\epsfbox{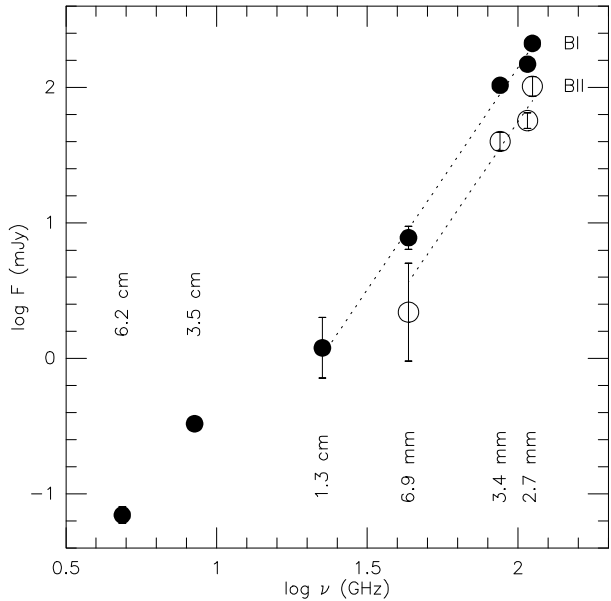} 
\caption{
Spectral energy distributions of the IRAS 4B sources.
Filled circles: BI.
Open circles: BII.
Flux densities are from Rodr{\'\i}guez et al. (1999), Reipurth et al. (2002),
Di Francesco et al. (2001), Choi et al. (2001), and Looney et al. (2000).
The flux uncertainties are smaller than the size of markers
except for the 1.3 cm and 6.9 mm data.
Dotted lines:
Best-fit power-law spectra in the (1.3 cm, 2.7 mm) section.}
\end{figure}

The mass of the circumstellar molecular gas can be derived
from the dust continuum in the millimeter bands,
using the procedure and equations described in Section 5 of Choi (2001)
and the mass emissivity given by Beckwith \& Sargent (1991).
The best-fit spectrum gives an opacity index of $\beta \approx 1.3$.
Assuming a distance of 235 pc to the source and a dust temperature of 33 K
(Hirota et al. 2008; Jennings et al. 1987),
the mass of IRAS 4BI is 1.0 $\pm$ 0.5 $M_\odot$,
which may include the inner protostellar envelope and the disk.
For comparison,
the mass estimates of BI in previous works (scaled to the distance of 235 pc)
are 0.43 $M_\odot$ in Looney et al. (2000), 1.1 $M_\odot$ in Choi (2001),
and 0.47/0.26 $M_\odot$ in J{\o}rgensen et al. (2007).
The mass of the whole protostellar envelope of BI
derived from single-dish observations
is $\sim$3.2 $M_\odot$ (Enoch et al. 2009, scaled to 235 pc).

\subsubsection{Outflow and Infrared Images}

IRAS 4BI drives a bipolar outflow roughly in the north-south direction.
Interferometric images of several molecular lines show
the redshifted lobe to the north
and the blueshifted lobe to the south of BI
(Figure 5; Choi 2001; Di Francesco et al. 2001;
J{\o}rgensen et al. 2007; Choi et al. 2011).
In these millimeter lines
the northern and southern lobes are often comparable in brightness.
Images of the infrared molecular hydrogen line
show a jet-like flow to the south only
(Hodapp \& Ladd 1995; Choi et al. 2006).
The bright H$_2$ knots are referred to as HL 9a/b (Figure 6a).
The H$_2$ jet may be bipolar in nature,
but the unseen northern jet
may be obscured by the dense protostellar envelope.
Recent far-IR observations showed highly excited H$_2$O emission
from the southern outflow (Herczeg et al. 2011, submitted to A\&A).

\begin{figure}[!t]
\centering \epsfxsize=8cm 
\epsfbox{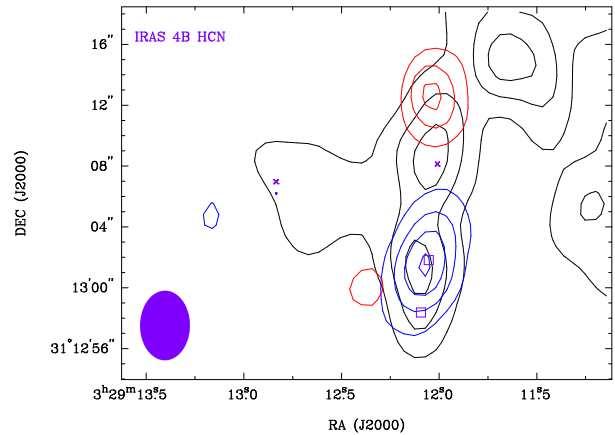} 
\caption{
Maps of the HCN $J$ = 1 $\rightarrow$ 0 line toward the IRAS 4B region,
from the data presented by Choi (2001).
For the line core of the $F$ = 2 $\rightarrow$ 1 hyperfine component
(black contours),
the HCN line was averaged
over the $V_{\rm LSR}$ interval of 5.8--7.2 km s$^{-1}$.
The lowest contour level and the contour interval are 150 mJy beam$^{-1}$,
and the rms noise is 50 mJy beam$^{-1}$.
For the blueshifted and redshifted outflows
(blue and red contours, respectively),
see Figures 1 and 2 of Choi (2001) for the velocity intervals.
The lowest contour level and the contour interval are 60 mJy beam$^{-1}$,
and the rms noise is 20 mJy beam$^{-1}$.
Shown at the bottom left-hand corner is the synthesized beam:
FWHM = 4\farcs6 $\times$ 3\farcs3 and PA = 0$^\circ$.
Crosses:
the 6.9 mm continuum sources.
Squares:
The H$_2$ line peaks HL 9a/b (Choi et al. 2006; Figure 6a).}
\end{figure}

\begin{figure*}[!t]
\centering
\epsfxsize=17cm 
\epsfbox{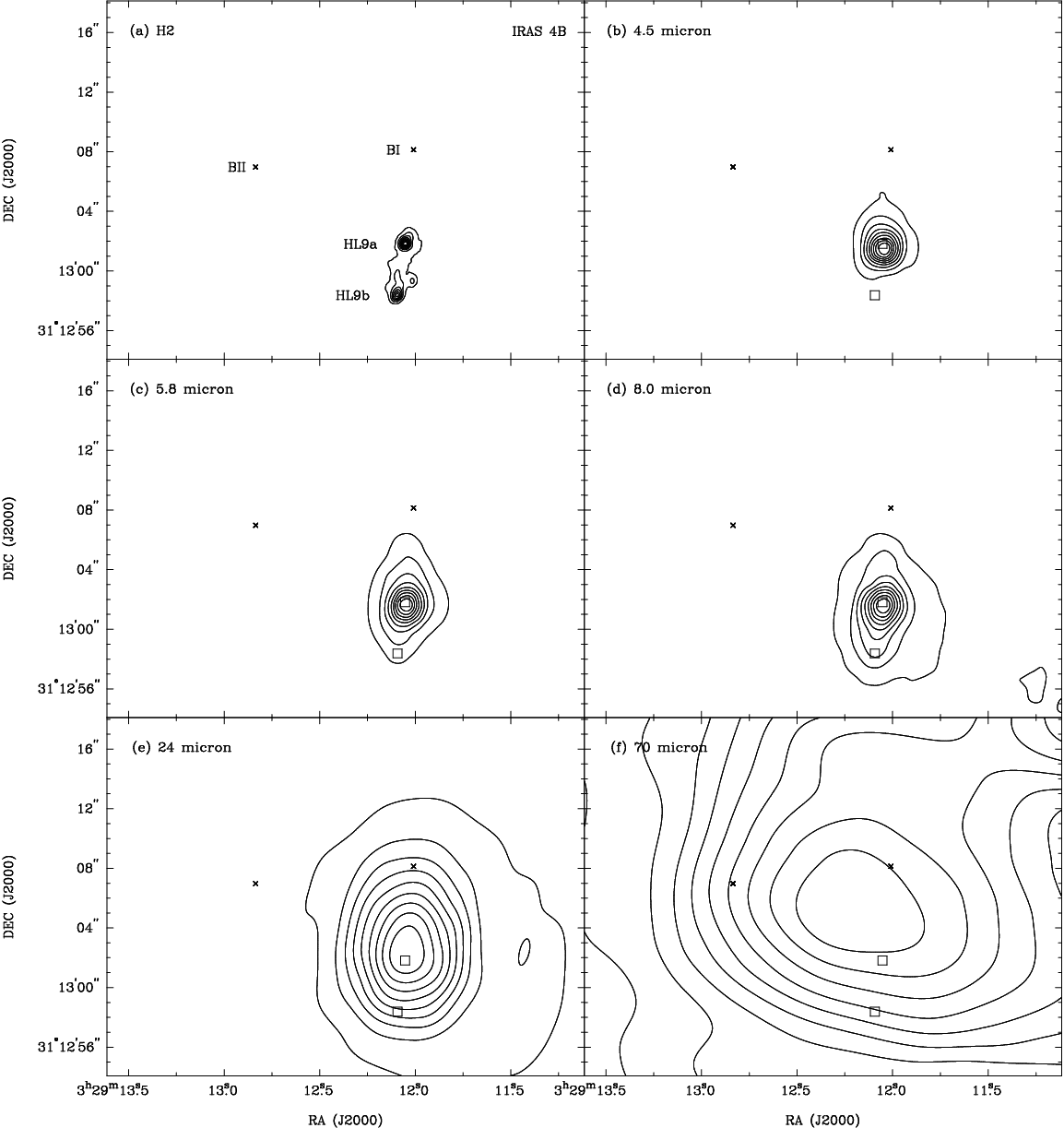}
\centerline{\scriptsize
[See http://minho.kasi.re.kr/Publications.html for the original figure.]}
\vspace{-\baselineskip}
\caption{
IR images of the IRAS 4B region.
(a) Map of the H$_2$ 1--0 $S$(1) line.
The emission peak positions are
$\alpha_{2000}$ = 03$^{\rm h}$29$^{\rm m}$12\fs05,
$\delta_{2000}$ = 31\arcdeg13$'$01\farcs8 for HL 9a
and 03$^{\rm h}$29$^{\rm m}$12\fs09, 31\arcdeg12$'$58\farcs4 for HL 9b.
(b) 4.5 $\mu$m image.
(c) 5.8 $\mu$m image.
(d) 8.0 $\mu$m image.
(e) 24 $\mu$m image.
(f) 70 $\mu$m image.
The lowest contour level and the contour interval
are 10\% of the maximum intensity in each image.
The highest contour is at the 90\% level.
The grayscale images are in a logarithmic scale and show low-level features.
The H$_2$ image is from the Subaru telescope
(see Figure 2 of Choi et al. 2006),
and the others are from the {\it Spitzer} data archive.}
\end{figure*}

The direction of the outflow,
i.e., the position angle of HL 9
with respect to the 6.9 mm continuum source of IRAS 4BI,
is $\sim$174$^\circ$.
By contrast, the position angle of the small-scale jet
traced by the H$_2$O maser
is $\sim$151$^\circ$ (Rodr{\'\i}guez et al. 2002; Park \& Choi 2007).
To explain the difference in the outflow directions,
it was proposed
that the jet is precessing at a high rate
(Marvel et al. 2008; Desmurs et al. 2009).
(Marvel et al. (2008) also suggested that there are probably two outflows,
i.e., BI may be a binary system.)
The position angles of HL 9a and 9b with respect to BI
agree within $\sim$1$^\circ$
and show no sign of precession to the proposed direction.

In the infrared images of the IRAS 4B region,
the H$_2$ jet, especially the bright knot HL 9a,
is the dominating emission feature.
In the {\it Spitzer} images,
only HL 9a/b can be seen in the 3.6--24 $\mu$m bands (Figure 6b-e),
and neither BI nor BII is clearly detectable.
Even at the 70 $\mu$m band,
the peak position is located between BI and HL 9a (Figure 6f),
and the interpretation of the image and the flux is ambiguous.
Note that the {\it Spitzer} source SSTc2dJ032912.04+311301.5
corresponds to HL 9a,
and the SED of this source (Per-emb 13) presented by Enoch et al. (2009)
is a combination of the flux densities of HL 9a on the short-wavelength side
and those of BI on the long-wavelength side.
Therefore, their bolometric temperature and luminosity of the BI protostar
may be overestimated.
This example suggests that, at least for some young stellar objects,
the emission from secondary sources
(such as outflows and/or binary companion)
can ``contaminate'' the infrared flux and make the SED appear hotter.
Such effects can introduce a bias
into SED surveys of young stellar objects.

\subsection{IRAS 4BII}

IRAS 4BII has been detected as a compact source
in interferometric images of millimeter and submillimeter continuum
(Looney et al. 2000; Choi 2001; Di Francesco et al. 2001;
J{\o}rgensen et al. 2007; J{\o}rgensen \& van Dishoeck 2010)
and in a single-dish submillimeter-continuum map (Smith et al. 2000).
However, its evolutionary status and star-formation activity
are not well known.
Single-dish radio observations usually do not have angular resolutions
high enough to separate BII from BI,
and BII has never been detected in infrared.
In molecular lines, BII was detected in the $^{13}$CO and HCN line maps only
(Figure 5; Choi 2001).
There is no known molecular outflow associated with BII.
The nondetection in the centimeter continuum
(Rodr{\'\i}guez et al. 1999; Reipurth et al. 2002)
also suggests that there is no detectable radio jet.
Since there is no detection in infrared,
the SED of BII is incomplete
and cannot be classified well enough to tell the evolutionary status.

The best-fit millimeter SED of IRAS 4BII (Figure 4)
gives a mass of $\sim$0.4 $M_\odot$,
using the same procedure and assumptions in Section 4.1.2.
The mass estimate, however, is very uncertain
because the 6.9 mm flux density (hence the opacity index)
has a relatively large uncertainty.
The uncertainty interval of the mass estimate is 0.1--2.3 $M_\odot$.
For comparison, the mass estimates of BII in previous works
(scaled to the distance of 235 pc)
are 0.12 $M_\odot$ in Looney et al. (2000), 2.2 $M_\odot$ in Choi (2001),
and 0.17/0.074 $M_\odot$ in J{\o}rgensen et al. (2007).
Note that the dust temperature of 33 K used in the mass estimate
came from the SED of the whole IRAS 4 system (Jennings et al. 1987).
Since BII is undetectable in infrared, it can be much colder,
and the mass can be larger than the values given above.

On the nature of IRAS 4BII, previous studies suggested
that it may be an optical/IR source (such as T Tauri star)
or a Class I young stellar object (Looney et al. 2000; Choi 2001).
The nondetection of BII in the {\it Spitzer} images (Figure 6), however,
seems to rule out these possibilities.
One possible explanation
(that was not known at the time of the previous studies)
is that it can be a newly recognized type of young stellar objects
known as very low luminosity objects
(VeLLOs; Young et al. 2004; Crapsi et al. 2005; Lee 2007; Dunham et al. 2008).
The internal luminosity of a VeLLO is less than 0.1 $L_\odot$,
which is much lower than what is predicted
by models of low-mass star formation (Shu et al. 1987).

Dunham et al. (2008) found
that the 70 $\mu$m flux correlates well with the internal luminosity.
Assuming that the strong source in the 70 $\mu$m image (Figure 6f)
is IRAS 4BI,
the image was fitted with a point source.
The residual map gives an upper limit of $\sim$500 mJy
on the flux density of BII.
Using the scaling given by Dunham et al. (2008),
the upper limit on the internal luminosity of BII is $\sim$0.08 $L_\odot$.
Therefore, it is possible that BII may be a VeLLO.

The compact structure of IRAS 4BII
and the nondetection of any star-formation activity
point to yet another possibility: the first hydrostatic core
(FHSC; Boss \& Yorke 1995; Masunaga et al. 1998).
Since the observational characteristics of FHSC
have not been well established yet (e.g., Enoch et al. 2010),
it is not clear if BII is such an object.
More sensitive and higher-resolution observations in the far-IR band
are needed to verify the status of BII
as VeLLO, FHSC, or any other possibilities.

\acknowledgments{
We thank Miju Kang and Gregory J. Herczeg
for helpful discussions and suggestions.
M. C. was supported by the Core Research Program
of the National Research Foundation of Korea (NRF)
funded by the Ministry of Education, Science and Technology (MEST)
of the Korean government (grant number 2011-0015816).
J.-E. L. was supported by the Basic Science Research Program
through NRF funded by MEST (grant number 2011-0004781).
The National Radio Astronomy Observatory is
a facility of the National Science Foundation
operated under cooperative agreement by Associated Universities, Inc.
This work is based in part on observations
made with the {\it Spitzer Space Telescope},
which is operated by the Jet Propulsion Laboratory,
California Institute of Technology, under a contract with NASA.}

\end{document}